\documentclass[manuscript,screen,sigconf]{acmart}

\usepackage{multirow}

\usepackage{graphicx}
\usepackage{xcolor}
\usepackage{soul}
\usepackage{wrapfig}
\usepackage{subcaption}
\usepackage{balance}
\newcommand{\ignore}[1]{}

\AtBeginDocument{%
  \providecommand\BibTeX{{%
    \normalfont B\kern-0.5em{\scshape i\kern-0.25em b}\kern-0.8em\TeX}}}

\copyrightyear{2024} 
\acmYear{2024} 
\setcopyright{rightsretained} 
\acmConference[CHI EA '24]{Extended Abstracts of the CHI Conference on
Human Factors in Computing Systems}{May 11--16, 2024}{Honolulu, HI, USA}
\acmBooktitle{Extended Abstracts of the CHI Conference on Human Factors in
Computing Systems (CHI EA '24), May 11--16, 2024, Honolulu, HI, USA}
\acmDOI{10.1145/3613905.3651057}
\acmISBN{979-8-4007-0331-7/24/05}
\begin{document}
\setlength{\tabcolsep}{1.5pt}
\title[Generative-AI in Content Creation]{A Preliminary Exploration of YouTubers' Use of Generative-AI in Content Creation}


\author{Yao Lyu}
\orcid{0000-0003-3962-4868}
\affiliation{%
  \institution{Pennsylvania State University}
  \city{University Park}
  \state{Pennsylvania}
  \country{USA}
}
\email{yaolyu@psu.edu}

\author{He Zhang}
\affiliation{%
  \institution{Pennsylvania State University}
  \city{University Park}
  \country{USA}}
\email{hpz5211@psu.edu}
\orcid{0000-0002-8169-1653}

\author{Shuo Niu}
\email{shniu@clarku.edu}
\orcid{https://orcid.org/0000-0002-8316-4785}
\affiliation{%
  \institution{Clark University}
  \streetaddress{950 Main St.}
  \city{Worcester}
  \state{MA}
  \country{USA}
  \postcode{01610}
}

\author{Jie Cai}
\orcid{0000-0002-0582-555X}
\affiliation{%
  \institution{Pennsylvania State University}
  \city{University Park}
  \state{Pennsylvania}
  \country{USA}
}
\email{jie.cai@psu.edu}
\renewcommand{\shortauthors}{Lyu et al.}

\begin{abstract}
Content creators increasingly utilize generative artificial intelligence (Gen-AI) on platforms such as YouTube, TikTok, Instagram, and various blogging sites to produce imaginative images, AI-generated videos, and articles using Large Language Models (LLMs). Despite its growing popularity, there remains an underexplored area concerning the specific domains where AI-generated content is being applied, and the methodologies content creators employ with Gen-AI tools during the creation process. This study initially explores this emerging area through a qualitative analysis of 68 YouTube videos demonstrating Gen-AI usage. Our research focuses on identifying the content domains, the variety of tools used, the activities performed, and the nature of the final products generated by Gen-AI in the context of user-generated content. 
\end{abstract}

\ccsdesc[500]{Human-centered computing~Empirical studies in collaborative and social computing}
\ccsdesc[500]{Human-centered computing~Empirical studies in HCI}

\keywords{Generative AI, YouTube, Content Creator, User-generated Content, Artificial Intelligence, Content Creation, Affiliated Marketing, Professional Development}

\maketitle

\section{Introduction}

The research and discussions surrounding generative artificial intelligence (Gen-AI) have been intense and ongoing in HCI (e.g., \cite{10.1145/3491101.3503719, 10.1145/3544549.3573794}). Gen-AI has been applied across a variety of domains, including machine learning \cite{foster2022generative,harshvardhan2020comprehensive}, image generation \cite{10.1145/3544549.3577043},  and text and audio processing \cite{10.1145/3511599, 10.1145/3411764.3445219,10.1145/3544548.3581402}. Owing to their creative capabilities, Gen-AI has been significantly adopted by content creators in fields such as art~\cite{galanter2016generative,10.1145/3544549.3577043}, education~\cite{article_1337500}, research~\cite{zhang2023redefining}, and entertainment~\cite{kumaran2019generating,10.1145/3411764.3445219,10.1145/3544548.3581402,singer2022makeavideo}. After over two decades of development, Gen-AI, with its high dependency on content, has emerged as one of the most impacted fields~\cite{KAPLAN201059}. Content creators have rapidly amplified the impact of Gen-AI through network dissemination, presenting both opportunities and challenges. On the positive side, Gen-AI applications have significantly enhanced content creation on social media by reducing the cost of content creation through user-friendly interactions and efficient performance and breaking down skill barriers, thus augmenting human creativity. Conversely, Gen-AI poses certain risks to communities, primarily because it is capable of creating fabricated content, leading to concerns about fake news~\cite{altay2023headlines, 10.1145/3411764.3445699} and the subtle misuse of misinformation~\cite{10.1145/3603163.3609064,li2023masquerade,charness2023generation,10.1145/3290605.3300469,10.1145/3544548.3580688}. This proliferation of AI content may overshadow high-value content, making it more challenging to find useful information and potentially contributing to the homogenization of content on social media~\cite{padmakumar2023does}.

In this context, conducting research on content related to Gen-AI on social media platforms can aid in understanding the current impact of Gen-AI on social media and the actual trends of Gen-AI-related content. This can facilitate future policy deployment, content analysis, and other welfare initiatives. Furthermore, as social media content producers, content creators play a crucial role in developing social media platforms \cite{bartolome2023literature,niu2023building}. Therefore, exploring these creators' thoughts and production intentions can help further understand Gen-AI's audience and its dissemination on social media. In this study, we primarily focus on videos and their creators on YouTube that generate content related to the theme of Gen-AI. Inspired by previous frameworks that investigate the roles of Gen-AI \cite{hwang2022too}, we propose the following research questions:




\begin{description}
   \item[RQ1] What domains do YouTubers explore with Gen-AI in the video?
   \item[RQ2] What types of Gen-AI tools do YouTubers use in the video?
   \item[RQ3] What actions do Gen-AI tools take in the videos?
   \item[RQ4] What Gen-AI products do the YouTubers primarily present in the video?
\end{description}


\section{Methods}
To gather videos demonstrating the use of Gen-AI, we employed a combination of search operators with the key phrase: \textit{``How to" (edit OR generate OR create OR make OR use) ``AI" (text OR image OR audio OR video OR animation OR content).} The search was constrained to videos published between January 1, 2023, and October 27, 2023, providing a data window of 300 days. This initial search yielded a total of 14,163 videos. To refine this dataset, we first programmatically excluded non-English videos. Next, we manually identified and applied 444 tags, added by YouTubers related to AI names or concepts, to filter the video selection further. This process resulted in a narrowed pool of 3,814 videos. From this refined group, we randomly selected 90 videos for in-depth analysis. After a review process to ensure relevance to our research questions, we were left with 68 videos. These videos were then utilized to develop a codebook and subsequent annotation.

\par
Initially, each researcher independently performed open coding on the videos, guided by the four research questions. After completing this step, we employed affinity diagramming on Miro to consolidate the independently coded content. Subsequently, all authors gathered to synchronize their individual coding efforts. Through collective agreement, we achieved consistent labeling for each video. We identified the connections between various concepts, leading to the systematic categorization and organization of these concepts. This process involved four cycles of the collective agreement, each focused on finalizing the codebook for one specific research question. As a result, we developed a unique codebook for each research question. In the final cycle, the researchers achieved substantial agreement for each research question, with Krippendorff's alpha scores of at least 0.796. In the concluding phase, pairs of researchers independently applied the developed codebook to code the data. Upon completing the coding, all researchers engaged in multiple joint meetings to exchange notes, revisit the original data, and discuss the outcomes of the coding process.





\section{Results}

\subsection{RQ1: What domains do YouTubers explore with Gen-AI in the video?}
\begin{figure*}[!ht]
    \centering
    \includegraphics[width=\textwidth]{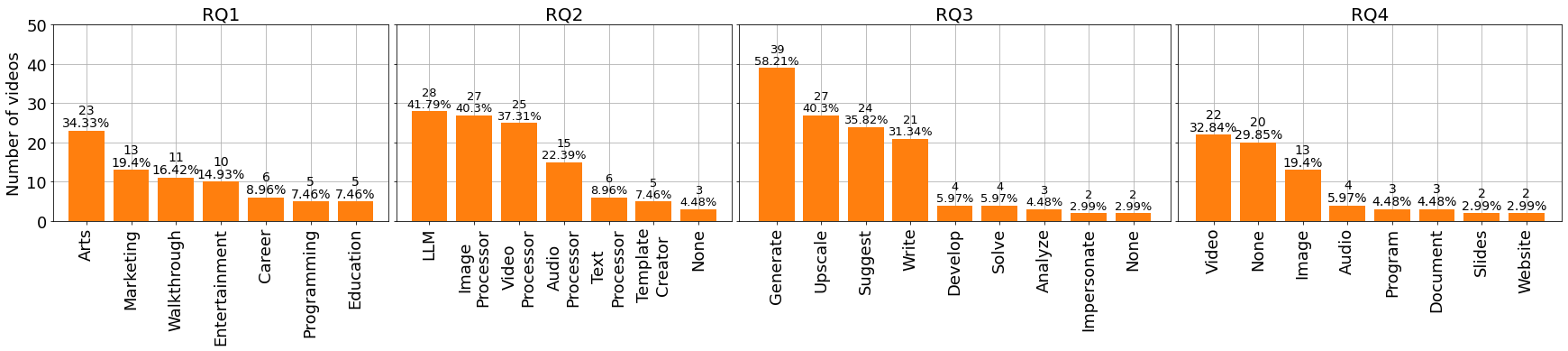}
    \caption{Distribution of videos across subcategories of the four research questions.}
    \label{fig:distribution}
     \Description{Bar charts to show the distribution of videos across subcategories of the four research questions.}
\end{figure*}

Overall, we identified six primary domains in which YouTubers have applied Gen-AI tools: \textit{marketing}, \textit{career}, \textit{arts}, \textit{entertainment}, \textit{programming}, and \textit{education} (see Figure-\ref{fig:distribution}, RQ1). Additionally, another category, \textit{walkthrough} (16.42\%), consists of videos that exclusively explain Gen-AI capabilities without focusing on a specific domain. These videos typically showcase Gen-AI tools or offer step-by-step tutorials.

\par

\begin{figure*}[!ht]
    \centering
    \includegraphics[width=1\textwidth]{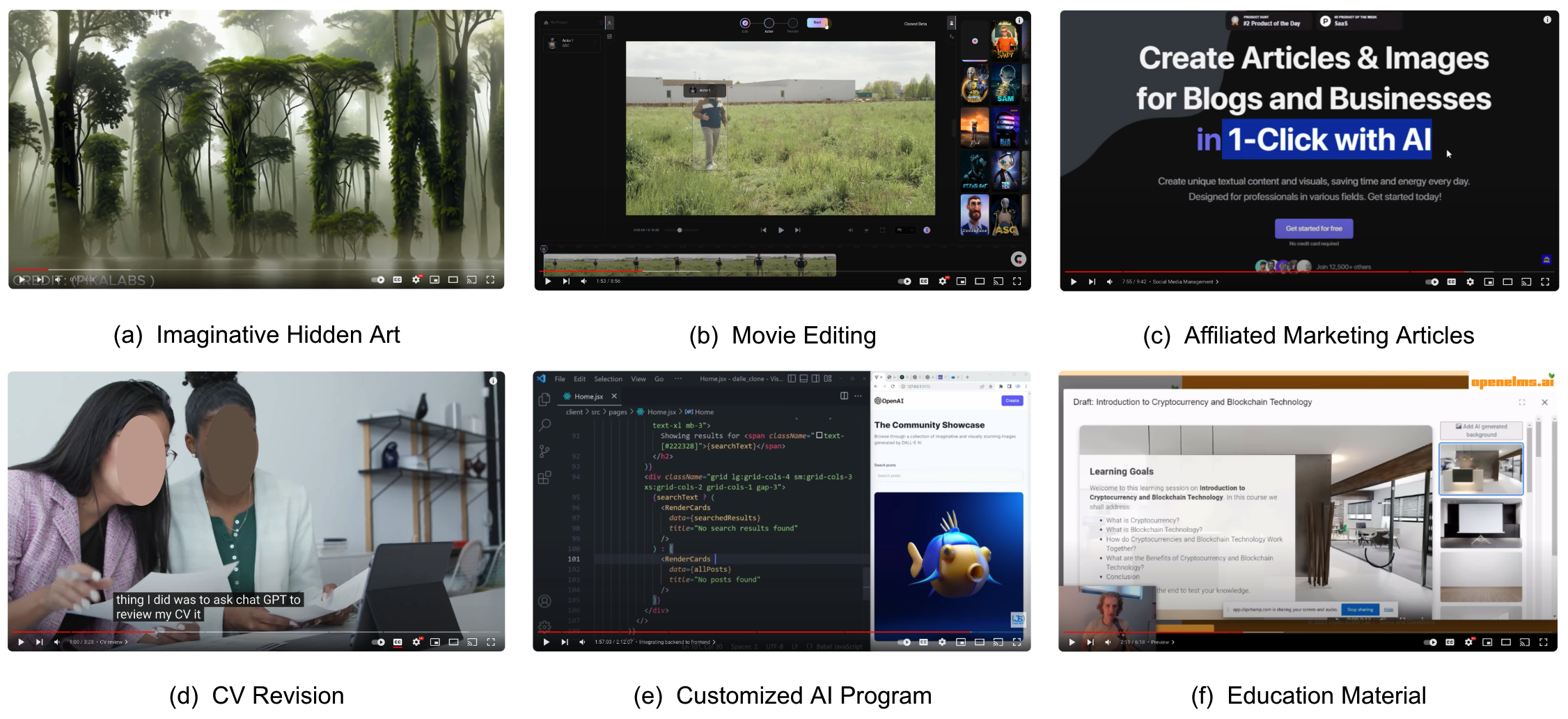}
    \caption{Examples of Domains}
    \label{fig:RQ1}
    \Description{Six YouTube screenshots to show examples of GenAI's applications in different domains: (a) Imaginative Hidden Art, (b) Movie Editing, (c) Affiliated Marketing Articles, (d) CV Revision, (e) customized AI Program, (f) Education Material}
\end{figure*}

\textbf{Arts} (34.33\%) refers to videos where Gen-AI is utilized to craft various digital art forms, such as AI-generated images and designs for graphic elements in posters, websites, and architectural projects. In the realm of digital arts, this typically involves the generation of artistic and imaginative imagery. For example, a YouTuber demonstrates using image-generating AIs to create a picture that cleverly incorporates a hidden word, as illustrated in Figure-\ref{fig:RQ1}a.
\par
\textbf{Marketing} (19.4\%) features videos where YouTubers use Gen-AI for affiliated marketing or business growth. This category includes tutorials and discussions on monetizing AI-generated content, leveraging AI to identify sales opportunities, and earning money through freelancing. For example, in a video talking about \textit{``How To Use AI For Affiliate Marketing''} (refer to Figure-\ref{fig:RQ1}c), the YouTuber demonstrates how to create a review video with ChatGPT and Pictory to earn commission from Amazon. Additionally, another video shows a YouTuber using Gen-AI to craft compelling product reviews to boost sales (Figure-\ref{fig:RQ1}c). 
\par
\textbf{Entertainment} (14.93\%) includes videos in which Gen-AIs are employed to generate various forms of entertainment media, such as movies, games, and music. A notable example, depicted in Figure-\ref{fig:RQ1}b, features a YouTuber using Gen-AI to automatically identify, select, create, and modify film characters, showcasing AI's transformative potential in entertainment.
\par
\textbf{Career} (8.96\%) is defined as videos that focus on using Gen-AI for professional development. YouTubers offer AI tool guidance to individuals seeking to improve expertise or increase productivity. The videos cover how to use tools like ChatGPT for tasks such as drafting resumes (illustrated in Figure-\ref{fig:RQ1}d), enhancing professional skills, boosting productivity, and managing finances. These videos cater to viewers who are interested in advancing their careers through the integration of AI technologies.
\par
In the \textbf{programming} domain (7.46\%), videos act as practical guides, showing how to apply Gen-AI technologies in various programming endeavors. For example, a prominent video offers a tutorial on using DALL·E to develop a customized AI image generator, exemplifying a practical application of AI in programming (illustrated in Figure-\ref{fig:RQ1}e). 
\par
\textbf{Education} (7.46\%) videos exemplify the adoption of Gen-AI for creating educational materials for classes or learners. These videos feature Gen-AI tools that assist in developing slides, quizzes, and other educational resources, as depicted in Figure-\ref{fig:RQ1}e.

\subsection{RQ2: What types of Gen-AI tools do YouTubers use in the video?}
The second question addresses the Gen-AI tools mentioned in the how-to videos. Three videos refer to using artificial intelligence in general but do not specify any particular tool. All other videos showcase at least one specific Gen-AI tool. It's important to note that while some tools offer multiple functions across different media types, we categorize them based on the function utilized in the video. In total, six types of tools are used in the videos (Figure-\ref{fig:distribution}, RQ2).
\par
The predominant tools are \textbf{LLMs} (41.79\%), with most powered by ChatGPT. A few videos use other LLMs, such as Palm and RapidAPI.
\par
YouTubers have showcased a diverse range of \textbf{image processors}, appearing in 40.30\% of the videos. The majority of these image Gen-AI tools, such as Midjourney, DALL·E, SeaArt, and Lexica, specialize in generating imaginative pictures on various topics from user-provided prompts. Tools like Picture Perfect AI, AutoPortrait.ai, and ReflectMe are designed for creating AI avatars. Additionally, image processors like Fotor and Adobe Firefly can enhance images intelligently.
\par
\textbf{Video processors} appear in 37.31\% of the videos. Like image processors, various Gen-AI tools are used for various video processing tasks. Some tools, such as Adobe Premiere Pro, Wisecut, and DScript, are designed for automated video editing or incorporate AI functionalities in video cutting. Another prevalent type includes tools like Pictory, Lumen5, and Steve.AI, which automatically generate videos or identify stock footage based on scripts or user searches. Recognizing that speaking directly to the camera is a popular method for creating YouTube content, many YouTubers have showcased tools like Synthesia, D-ID, and BHuman for generating talking avatar videos. Tools like Kaiber are also used to animate AI-generated images and integrate them into video creation.
\par
AI tools used for \textbf{audio processing} are in 22.39\% of the videos. Most of these are text-to-speech tools like Voxbox, Play.ht, and ElevenLabs. Additionally, there are tools dedicated to supporting podcast editing, like AudioPod.
\par
Gen-AI tools for \textbf{text processing} are featured in 8.96\% of the videos. Unlike LLMs, these Gen-AI tools are specifically designed to create well-formatted textual content for various scenarios. For instance, SEOWriting.AI and Zentask can generate content optimized for website search engine optimization (SEO). VidIQ and VEED.IO can craft popular video scripts, aiding in producing engaging video content. Quillbot and Rytr are tools used for paraphrasing or creative writing.
\par
\textbf{Template creator} tools (7.46\%) are Gen-AIs that assist in creating materials like slides or website templates. Examples of such tools include Canva Magic Write and Openelms.ai for slide creation and Durable.co for web templates.

\begin{table*}[!ht]
    \caption{Types and Examples of Gen-AI Tools}
    \centering
    \scalebox{0.8}{
    \begin{tabular}{|p{0.2\textwidth}|p{1\textwidth}|}
    \hline
    Tool Type & Example Applications\\
    \hline
    Large Language Models & ChatGPT (\url{https://chat.openai.com/}), Palm (\url{https://ai.google/discover/palm2/}), RapidAPI (\url{https://rapidapi.com/})  \\
    \hline
    Image Processor &  Midjourney (\url{https://www.midjourney.com/}), DALL·E (\url{https://openai.com/dall-e-2}), SeaArt (\url{https://www.seaart.ai/}), Lexica (\url{https://lexica.art/}), Picture Perfect AI (\url{https://pictureperfect.ai/}), AutoPortrait.ai, (\url{https://autoportrait.ai/}), ReflectMe (\url{https://reflectme.art/}), Fotor (\url{https://www.fotor.com/}), Adobe Firefly (\url{https://www.adobe.com/products/firefly.html}) \\
    \hline
    Video Processor &  Adobe Premiere Pro (\url{https://www.adobe.com/products/premiere.html}), Wisecut (\url{https://www.wisecut.video/}), DSript (\url{https://www.descript.com/}), Pictory (\url{https://pictory.ai/}), Lumen5 (\url{https://lumen5.com/}), Steve.AI (\url{https://www.steve.ai/}), Sythesia (\url{https://www.synthesia.io/}), D-ID (\url{https://www.d-id.com/}), BHuman (\url{https://app.bhuman.ai/}), Kaiber (\url{https://kaiber.ai/}) \\
    \hline
    Audio Processor &  VoxBox (\url{https://filme.imyfone.com/voice-recorder/}), Play.ht (\url{https://play.ht/}), ElevenLabs (\url{https://elevenlabs.io/}), AudioPod (\url{https://audiopod.cloud/}) \\
    \hline
    Text Processor &  SEOWriting.AI (\url{https://seowriting.ai/}), ZenTask (\url{https://zentask.ai/}), VidIQ (\url{https://vidiq.com/}), VEED.IO (\url{https://www.veed.io/}), Quillbot (\url{https://quillbot.com/}), Rytr (\url{https://rytr.me/}) \\
    \hline
    Template Creator&  Canva Magic Write (\url{https://www.canva.com/magic-write/}), Openelms.ai (\url{https://www.openelms.ai/}), Durable.co (\url{https://durable.co/}) \\
    \hline
    \end{tabular}
    }
    \label{tab:my_label}
\end{table*}


\subsection{RQ3: What actions do Gen-AI tools take in the videos?}
We also recognized eight types of actions that Gen-AI tools take in the YouTube videos, including \textit{generate}, \textit{upscale}, \textit{suggest}, \textit{write}, \textit{develop}, \textit{solve}, \textit{analyze}, and \textit{impersonate} (Figure-\ref{fig:distribution}, RQ3).

\begin{figure*}[!ht]
    \centering
    \includegraphics[width=1\textwidth]{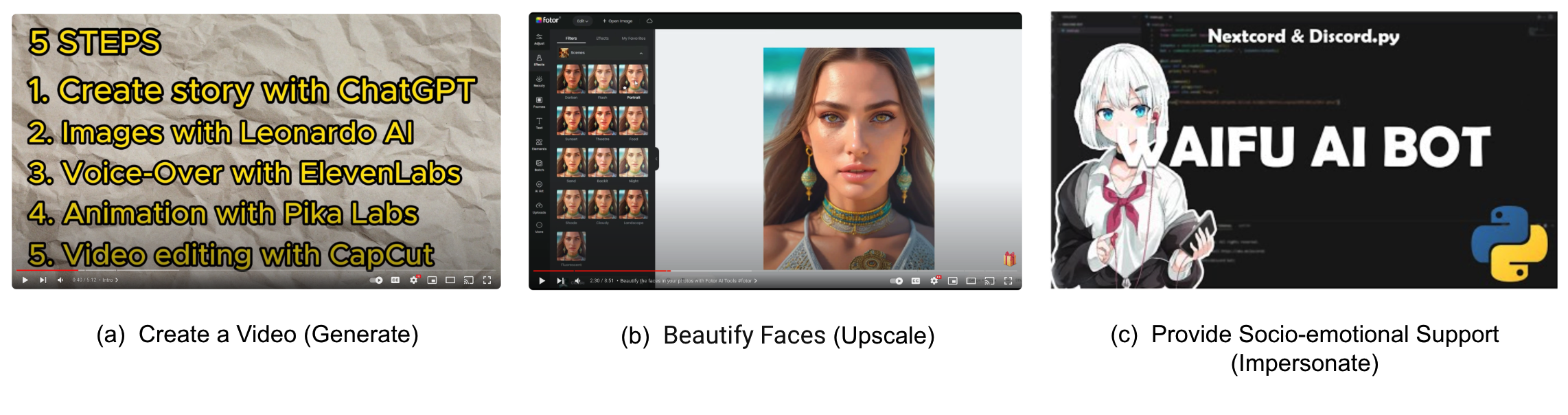}
    \caption{Examples of Actions}
    \label{fig:RQ3}
     \Description{Three YouTube screenshots to the example of actions: (a) Create a Video (Generate), (b) Beautify Faces (Upscale), (c) Provide Socio-emotional Support (Impersonate).}
\end{figure*}

\textbf{Generate} involves YouTubers using Gen-AI to automatically produce new content, including texts, images, videos, and audio, by utilizing prompts or converting one media type into another. A majority of these videos (58.21\%) utilize Gen-AI for content generation. This highlights that obtaining AI-generated material is the primary application of Gen-AI tools among YouTube creators. In arts and marketing videos, creators demonstrate to viewers how to create AI-generated images or videos. There is a notable interest in tutorials instructing users on effectively utilizing and refining AI prompts for image generation. In video creation tutorials, the materials most frequently generated by YouTubers include video storylines, images or stock footage for video clips, and AI-generated voice-overs. For example, one application showcased by a YouTuber involves the entire process of generating video materials using various Gen-AI tools -- ranging from story development and image/anime generation to voice-over synthesis and video editing, as illustrated in Figure-\ref{fig:RQ3}a. In educational videos, YouTubers also demonstrate the use of Gen-AI tools to create slides for classes.

\par
In contrast to "generate," \textbf{upscale} (40.30\%) involves the application of Gen-AI to existing content, to enhance it through AI effects or incorporate AI-generated elements. Particularly in videos focused on creating artistic material, these creators employ AI tools to improve the quality of their images or videos. Typical applications include using AI filters to beautify figures (as shown in Figure-\ref{fig:RQ3}b) or to perform artistic transformations of photographs. In entertainment, YouTubers often demonstrate how they use Gen-AIs to add special effects or virtual elements to their videos, further enhancing viewer engagement and visual appeal (refer to Figure-\ref{fig:RQ3}b).

\par
Beyond generating or upscaling digital content, YouTubers employ Gen-AIs for \textbf{suggesting} ideas or topics, observed in 35.82\% of the videos. Suggesting refers to providing multiple general ideas that are broadly relevant and might be useful to users' questions. This often involves using tools like ChatGPT to explore open-ended questions or to seek examples, suggestions, or recommendations. In arts-related content, creators use ChatGPT to generate prompt suggestions; for instance, a YouTuber in a digital art tutorial asked, \textit{``Give me some ideas for a Spider-Man costume,''} receiving a variety of prompts in response. In marketing, YouTubers turn to ChatGPT for insights into trending topics to inspire their content creation. One video tutorial on blogging for Medium.com, for example, starts with the creator asking ChatGPT, \textit{``What are the top 10 books for business?''} using the AI's suggestions to craft a book review blog. Similarly, in a guide on monetizing AI-generated short videos, the YouTuber prompts ChatGPT with, \textit{``Give me 6 philosophy lessons about self-discipline by Socrates,''} to brainstorm relevant video topics. In the educational domain, Gen-AIs serve as a tool for curating knowledge and formulating questions for students. A notable example includes a YouTuber using ChatGPT to \textit{``Create 10 multiple-choice questions based off of the following YouTube Video transcript.''} Career-focused videos also showcase the use of Gen-AIs for professional development; one YouTuber, for instance, shares his experience in a video talking about \textit{Asking An AI How To Get Video Production Clients''}
\par

Gen-AI tools are employed for \textbf{writing} content such as articles and video scripts, which appeared in 31.34\% of the videos. Write, in this sense, particularly refers to actions that produce well-formatted text documents ready to publish, such as articles and video scripts (in contrast to previously mentioned actions that generate raw text, e.g., ideas or recommendations). This category includes writing articles and various editing-related tasks, like summarizing and rephrasing texts. In the marketing and entertainment domains, YouTubers have adopted Gen-AIs to streamline the creation of video scripts and blog posts, thereby facilitating content monetization. An example of a typical prompt is: \textit{``Provide an outline for a YouTube video script titled: Essential Survival Skills: How to Survive A Bear Attack."} In educational settings, Gen-AI assists with the composition of essays and the development of slide content.

\par

In addition to the primary uses mentioned above, YouTubers engage in various other activities with Gen-AI tools, though these appear in fewer than 10\% of their videos. One such activity involves YouTubers \textbf{developing} applications using Gen-AI APIs, which they incorporate into programming tutorial videos. Additionally, some YouTubers demonstrate the capability of Gen-AI to \textbf{solve}. In contrast to the previously mentioned "suggesting," solving here means providing detailed and specific solutions to practical problems, such as answering complex math equations or addressing coding challenges. Another application seen in these videos is using Gen-AI to \textbf{analyze}; this action means using Gen-AI like ChatGPT to critically distinguish different subjects, ranging from book comparisons to evaluating the pros and cons between traditional and IR saunas. There are also instances where YouTubers have requested ChatGPT to \textbf{impersonate} specific characters for interactive segments in their videos. This action foregrounds Gen-AI's functionality of talking like an actual human and providing socio-emotional support to users ( Figure-\ref{fig:RQ3}c).

\subsection{RQ4: What Gen-AI products do the YouTubers primarily present in the video?}
To clarify the variety of outcomes YouTubers achieve with Gen-AI, we have categorized the array of final products featured in their videos (Figure-\ref{fig:distribution}, RQ4). Predominantly, when showcasing Gen-AI tools, YouTubers present the \textbf{videos} (32.84\%) created by these tools. They also display products in diverse formats, including \textbf{images} (19.30\%), \textbf{audios} (5.97\%), \textbf{computer programs} (4.48\%), \textbf{documents} (4.48\%) like blogs or write-ups, \textbf{slides} (2.99\%), and \textbf{websites} (2.99\%). Furthermore, 29.85\% of the videos do not include a presentation that introduces or demonstrates a specific product.

\section{Discussion}

There has been rich work on envisioning how HCI and Gen-AI can mutually contribute to each other \cite{10.1145/3491101.3503719,10.1145/3544549.3573794,weisz2022hai,fui2023generative,morris2023design,huang2023future}. Based on the domains, tools, actions, and outputs that YouTubers show in the videos, we also discuss the implications for the HCI community in this section. The implications target stakeholders related to content creation with Gen-AI. Specifically, we point out directions to designers so that they can better understand and facilitate content creation with Gen-AI. We also caution scholars in the HCI community about the problems that Gen-AI might cause.

\subsection{Gen-AI influencing Content Creation}
Our findings reveal that YouTubers increasingly leverage Gen-AI tools for artistic creation and entertainment content production. This trend is evident as approximately half of the analyzed videos pertain to the arts and entertainment categories. In our RQ2 analysis, it was observed that YouTubers utilize a variety of media processors for content creation within the realms of art and entertainment. Furthermore, the insights from RQ3 provide a deeper understanding of how YouTubers use Gen-AI to generate new content and upscale existing material. The discoveries in RQ4 highlight that YouTubers have also showcased AI-generated images, demonstrating the diverse applications of Gen-AI in their content creation process.
\par
The results have several implications for Human-Computer Interaction. First, for art creators, the influence of Gen-AI tools is significant. These AI-powered tools can simplify the content creation process by allowing creators to obtain ideas and generate materials. YouTubers are actively using ChatGPT to search for video topics. Platforms like Midjourney enable the generation of images. However, this ease of use may lead to a dependency on AI tools, prompting crucial questions about human agency in the creative process within arts and entertainment. This raises intriguing queries: Does content produced with Gen-AI assistance fully capture the creator's originality? And could reliance on Gen-AI potentially undermine human creativity over time? These concerns present fertile ground for future HCI research, particularly regarding the balance and interaction between human-AI collaboration and individual creative agency. Furthermore, the findings reveal that YouTubers often employ multiple Gen-AI tools for various aspects of video creation, such as scripting, image creation, voice generation, animation, and video editing. This emphasizes the variety of Gen-AI outputs \cite{weisz2023toward} and highlights a potential area for future research in developing integrated platforms that streamline the Gen-AI content creation process.
\par
Secondly, from the perspective of User-Generated Content platforms, the growing prevalence of AI-generated content (AIGC) necessitates a focus on developing methods to assess its quality. The ease of creation associated with Gen-AI tools suggests that their outputs might not match the quality of human-made content. This potential disparity in quality could lead to a decrease in viewer interest. Consequently, the surge in AIGC presents a challenge for maintaining user engagement on platforms dedicated to UGC. Future HCI research in platform governance could play a crucial role in devising or enhancing methods for evaluating the quality of AIGC, aiming to bolster its appeal and engagement among users.

\subsection{Gen-AI influencing Affiliated Marketing}
While prior work pointed out the use of Gen-AI for business \cite{houde2020business}, our findings specifically show that Gen-AIs offer new pathways for content creators to monetize their work and manage their business. Our analysis reveals that approximately 20 percent of the videos address existing challenges in the marketing sector and explore solutions utilizing Gen-AI tools. As highlighted in the response to RQ2, diverse tools are available for creating marketing content, including review articles and advertising videos. The findings from RQ3 delve into specific contributions of Gen-AI in marketing, such as offering inspiration for business strategies. Furthermore, the demonstrations in RQ4 provide a clearer understanding of how these products can be effectively used in marketing contexts. This comprehensive overview underscores the significant role of Gen-AI in revolutionizing marketing approaches and enhancing business opportunities for content creators.
\par
Firstly, our findings highlight the potential impact of Gen-AI on affiliate marketing. Many content creators produce videos for monetary gain or as a career. Gen-AI can assist in some tasks at a significantly reduced cost. Content creators can employ ChatGPT to brainstorm ideas, suggest information sources, and write articles. Furthermore, tools like Midjourney can be utilized to create images for products and design websites to promote business. Gen-AI's capabilities make it a knowledgeable, cost-effective, and easily accessible advisor. This development opens new avenues for future HCI research to investigate how content creators leverage Gen-AI, supporting emerging business models and affiliate marketing strategies.
\par
Secondly, in monetization programs, Gen-AI tools like ZenTask or ChatGPT allow content creators to produce product reviews efficiently with minimal effort. The simplicity and cost-effectiveness of Gen-AI tools could lead to an influx of AI-generated content in affiliate marketing. However, the monetization algorithms on most platforms are primarily quantity-focused, often giving precedence to the number of featured products and their view counts when evaluating a content creator's contributions. This trend of increasing Gen-AI-generated content may potentially distort these algorithmic assessments. Therefore, there is a significant opportunity for the HCI and recommendation algorithm communities to investigate and develop methods to adapt these algorithms. The goal would be to more effectively manage the monetization of affiliate marketing content, ensuring a fair and precise evaluation of creator contributions in this rapidly changing environment.

\subsection{Gen-AI Influencing Professionalism Development}
Lastly, Gen-AI is set to significantly revolutionize the areas of skill sharing and professional development, which are fundamental to the culture of the YouTube community \cite{PreeceWeTube}. According to the data, about 25\% of the videos demonstrate the use of Gen-AI tools for educational purposes and individual professional advancement. Moreover, 16.42\% of the videos, though not dedicated to a specific domain, offer comprehensive tutorials on applying Gen-AI tools. This indicates the widespread use of Gen-AI in the YouTube community: facilitating learning and aiding YouTubers in improving their skills. The insights from RQ3 also shed light on the varied functions of Gen-AI in education and training, helping users more efficiently address challenges in their professional pursuits.
\par
While the findings echo previous investigations on how Gen-AIs increase task productivity \cite{muller2022drinking}, they also suggest several promising directions for future research in Human-Computer Interaction. One key area could involve investigating the challenges encountered by creators with limited video creation skills and those belonging to disadvantaged communities. The skills discussed in the findings frequently require engaging with content across various modalities. For instance, video creation necessitates combining text, images, and audio, which might pose accessibility issues for people with visual or hearing impairments. Furthermore, there is unequal access to Gen-AI tools among different creator communities, potentially leading to social justice concerns on video-sharing platforms \cite{lyu2023because,lyu2024got}. Previous work has emphasized the importance of designing Gen-AI for co-creation \cite{weisz2024design}. Based on our findings, researchers should examine whether and how Gen-AI technologies are inclusive and benefit content creators from various backgrounds, ensuring equity and fairness in their use and impact. Future studies can use qualitative studies, such as ethnography studies, to capture the situated and contextual nature of the use of Gen-AI by different populations, especially minority groups. Gen-AI developers can also employ participatory design methods to incorporate different user groups into developing Gen-AI tools. These studies should aim to bridge the digital divide caused by the emerging Gen-AI tools and create a socially fair environment for all content creators.



\balance

\bibliographystyle{ACM-Reference-Format}
\bibliography{references.bib,references1.bib}

\end{document}